# Multiresolution Analysis Techniques to Isolate, Detect and Characterize Morphologically Diverse Features of Structured ICF Capsule Implosions

Bedros Afeyan, Marine Mardirian, Peter Jones, Jean Luc Starck and Mark Herrmann

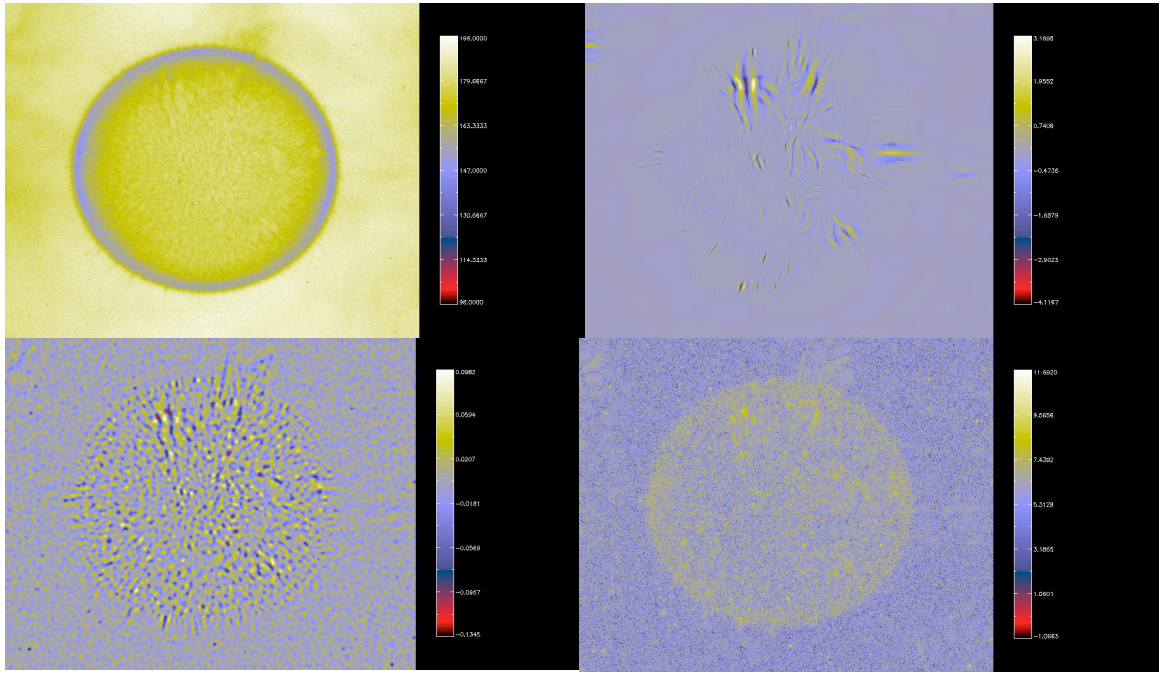

Fig. 1. Monochromatic 6.15keV X ray backlighting generated radiograph of a double Z pinch 2mm-scale hohlraum target, Z1561, the background removed wavelet-isolated image of intermediate scale correlated features (below it), its curvelet-detected features (to its right) and its WaSP function in log scale below the curvelet-detected features image.

*Abstract* – **In order to capture just how nonuniform and degraded the symmetry may become of an imploding inertial confinement fusion capsule one may resort to the analysis of high energy X ray point projection backlighting generated radiographs. Here we show new results for such images by using methods of modern harmonic analysis which involve different families of wavelets, curvelets and WaSP (wavelet square partition) functions from geometric measure theory. Three different methods of isolating morphologically diverse features are suggested together with statistical means of quantifying their content for the purposes of comparing the same implosion at different times, to simulations and to different implosion images.**

It is well known that target surface imperfections, radiation asymmetry and seeded hydrodynamic instabilities can cause an imploding ICF shell to form large modulations and distortions so that the assembled core, at the end of such an implosion, will never sustain propagating burn or alpha particle bootstrapping and will, at best, fizzle. To assess shell breakup or gross deformations, it is advantageous to have a sequence of images along the way and to compare them to integrated radiation-magnetohydrodynamic simulations to see if the growth of such early structures is predictable or whether they lie outside the scope of what is treated correctly in these complex multiphysics models. How small do these perturbations and imperfections have to be to be captured by radiographic techniques with sufficient numbers of high energy X ray photons is a compelling question, especially in

Manuscript received 1 December 2010; revised
B. Afeyan and M. Mardirian are with Polymath Research, Pleasanton, CA, P. Jones is at Yale University, J-L Starck is at CEA, France, M. Herrmann is at Sandia National Laboratories in Albuquerque, NM.
Work was supported by Sandia National Laboratories.
Publisher Identifier S XXXX-XXXXXXX-X

the presence of noise and other recording artifacts inherent in real data acquisition systems and clutter from large scale coarse features in the images which have nothing to do with the dense structures we wish to detect. The more sophisticated the analysis tools used are, the more leeway will be granted to the hardware and its idiosyncrasies. We have devised three such promising methods of analysis which can identify and separate physically interesting correlated features by first removing the large scale background or clutter features and then being left at the end with a texture resembling multifractal or multiscale, highly oscillatory, statistically almost invariant features which are characterized separately. In between these two, we will also have isolated highly correlated features which are on intermediate scales and which are most likely to be the culprits degrading the implosion qualities most directly.

The techniques are to use the MAD (median absolute deviation) technique on biorthogonal wavelet decomposed images [1,2], and peal off the large scale features and designate them as background. The isotropic, undecimated wavelet transform applied to the residue decomposes the image to 6 scales. Level 3, renders a very good image of the intermediate scale structures. This constitutes one isolation and detection method for that structure which has also proven useful in 6 other such images already analyzed. A second method involves taking the residue, ie Original image – background image, and applying a Curvelet transform [1,2,3] to it and with a five sigma thresholding scheme [1], reconstructing the features which are most dominant in a curve like, axially or azimuthally correlated way. This method has also proved useful in 6 other images from similar Z campaigns. The residue of this operation, ie original – background – curvelet detected, renders a texture like image which contains very little correlated, spatially extended structures and contains instead the highly fluctuating small scale features which tend to be spread throughout the target interior. This is left for separate statistical analysis. The third technique is to take the background subtracted image and calculate the WaSP function for it. The Wavelet Square Partition function is a low dimensional example of the more general Beta-number concept of Peter Jones which allows one to calculate length (or area in our case of a 2D image, and volume more generally) in a strict mutiscale fashion which is well suited for fractal or multifractal structure characterization [4]. Wherever an image contains structures on many scales this will lead to a large WaSP function value locally. Conversely, wherever features are smooth and concentrated on some specific scale, this will lead to a small WaSP function value, locally. Thus WaSP functions show us where there is a lot of correlated structure on different scales and where there is not. So this does not isolate or allow a quantification of the X ray transmission at those locations, but points out where there are such multiscale modulations. We used undecimated isotropic wavelet transforms in 2D for the WaSP function analysis.

The results of these three techniques are shown in the figure for the Sandia Z pinch machine shot Z1561 with the monochromatic 6.15 keV backlighting radiographic technique of very dense plasmas from ICF implosions [5]. The original image is also shown in [5]. We show the background removed levels 3 isolated image below the original as well as the curevlet detected features in the no background image next to the original, as well as the WaSP function showing where the multiscale, correlated features are. underneath the curvelet isolated mage. We have also performed these same analyses on the unwrapped images which accentuate edge features and clip off the central portions of the image. That results in a Euclidian (r, phi) representation. These serve a complimentary purpose, better detecting the statistics of the outer edge features by ignoring inner structures. The full multiresolution analysis of Z1561 and Z1562, which is a similar radiograph taken later in time after further compression, will be published elsewhere [6]. The statistical characterization of the textured residues, and the statistical analysis of the extended modulations on intermediate scales treated with various different techniques can be found in that paper as well. Such analyses should be useful for the comprehensive analysis of future Z and NIF radiographic images as long as there are hundreds of pixels per dimension. From radiation asymmetry to turbulence mix and all features in between should be detectable using these advanced techniques of harmonic analysis.